# RL2: Reinforce Large Language Model to Assist Safe Reinforcement Learning for Energy Management of Active Distribution Networks

Xu Yang, *Graduate Student Member, IEEE*, Chenhui Lin, *Member, IEEE*, Haotian Liu, *Member, IEEE*, and Wenchuan Wu, *Fellow, IEEE*

*Abstract*—As large-scale distributed energy resources are integrated into the active distribution networks (ADNs), effective energy management in ADNs becomes increasingly prominent compared to traditional distribution networks. Although advanced reinforcement learning (RL) methods, which alleviate the burden of complicated modelling and optimization, have greatly improved the efficiency of energy management in ADNs, safety becomes a critical concern for RL applications in real-world problems. Since the design and adjustment of penalty functions, which correspond to operational safety constraints, requires extensive domain knowledge in RL and power system operation, the emerging ADN operators call for a more flexible and customized approach to address the penalty functions so that the operational safety and efficiency can be further enhanced. Empowered with strong comprehension, reasoning, and in-context learning capabilities, large language models (LLMs) provide a promising way to assist safe RL for energy management in ADNs. In this paper, we introduce the LLM to comprehend operational safety requirements in ADNs and generate corresponding penalty functions. In addition, we propose an RL2 mechanism to refine the generated functions iteratively and adaptively through multi-round dialogues, in which the LLM agent adjusts the functions' pattern and parameters based on training and test performance of the downstream RL agent. The proposed method significantly reduces the intervention of the ADN operators. Comprehensive test results demonstrate the effectiveness of the proposed method.

*Index Terms*—Active distribution network, energy management, reinforcement learning, large language model, penalty function.

## I. INTRODUCTION

As significant distributed energy resources (DERs), such as diesel generators (DGs), photovoltaics (PVs), and battery energy storage systems (BESSs), are integrated into the distribution networks, traditional distribution networks are gradually transforming into active distribution networks (ADNs) [1], [2], which also involves massive emerging ADN operators. In order to fully exploit the potential of these DERs, energy management, which adjusts the active/reactive power generation of controllable devices [3], [4], is essential compared to passive control strategies. Effective energy management not only reduces operational costs but also improves the safety of the ADNs.

Traditionally, energy management is formulated as a coordinated active/reactive power optimization problem [5], [6], which can be successfully solved by existing well-developed optimization algorithms. However, these methods involve complicated modelling and optimization of the entire ADN and the controllable devices within it, which is usually unaffordable for the ADN operators. In addition, solving the optimization problem involves a heavy communication and computational burden, which makes it difficult to cope with the fluctuations in loads and PV power generation [7], [8]. Therefore, reinforcement learning (RL) based energy management methods recently received considerable attention [9]-[13]. Unlike the above optimization algorithms, RL-based methods construct the energy management problem as a Markov decision process (MDP) [14], in which the RL agent interacts with the environment and automatically updates its control policy based on the feedback reward signals. After the offline training, a trained RL agent is able to generate control strategies of controllable devices using only ADN online measurements and without continuously calculating optimization problems, which greatly improves the efficiency for ADN operators.

As can be seen from the definition of RL, the performance of the RL agent depends heavily on the feedback reward signal from the MDP, which requires careful design of the reward function. Fortunately, the design of the reward function is relatively straightforward for the energy management problem, which aligns with the objective function, i.e., minimizing operational costs of the ADN. When applying RL-based methods to real-world problems, safety becomes a critical concern. As for the operational safety constraints in ADNs, a common practice is to include a penalty term in the reward, providing the RL agent with appropriate feedback when a violation occurs. To meet the operational safety requirements in ADNs, researchers have developed several types of penalty functions, such as violation times, L1-norm function, L2-norm

This work was supported in part by the National Key Research and Development Plan of China under Grant 2022YFB2402900 and National Science Foundation of China under Grant U2066601. *(Corresponding author: Wenchuan Wu).*

The authors are with the State Key Laboratory of Power Systems, Department of Electrical Engineering, Tsinghua University, Beijing 100084, China (e-mail: yangxuthu@163.com, linchenhui@tsinghua.edu.cn, liuhaotian@tsinghua.edu.cn, wuwench@tsinghua.edu.cn).

function [15], bowl-shape function [16], and log-barrier function [17]. In addition, adaptively adjusting the parameters (especially the weight coefficients) of the penalty functions throughout the RL training process is also crucial for safety considerations. Therefore, some advanced constrained RL algorithms, such as constrained policy optimization [18] and Lagrangian method [19], have also been widely studied and applied in power system operation [20]-[23]. Although these methods have been shown to be effective in some cases, the design, testing, and tuning of the penalty functions still require extensive domain knowledge in RL and power system operation, which is costly in terms of time and expenses for these new ADN operators. A more flexible and customized approach to deal with these penalty functions is urgently needed by the ADN operators.

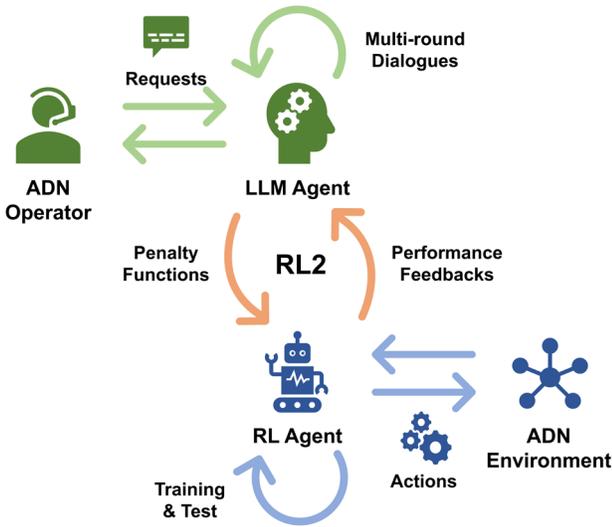

**Fig. 1.** Scheme of the proposed method.

This paper is inspired by the recent advancement of generative pre-trained transformer (GPT) based large language models (LLMs) in the fields of natural language processing, computer vision, and multimodal applications. Trained on a vast corpus of data from the Internet, these LLMs are embedded with a huge amount of knowledge and information, including the domain knowledge of RL and power system operation [24]. Moreover, the well-designed transformer structure and billions of parameters endow the LLMs with powerful capabilities in comprehension, reasoning, and in-context learning, which provide a novel and promising way to bridge the gap between the operator requirements and power system operation. Some pioneering studies have already attempted to apply LLMs in the power system, such as forecasting [25], document analysis [26], and simulation setup [27], [28]. Although the LLMs show great potential in various fields, LLMs by themselves cannot directly generate a safe and reliable control policy for energy management in ADNs. Considering the limitations of RL methods and existing LLMs, the main focus of this paper is on:

*How to combine the advantages of RL and LLMs to achieve safe energy management in ADNs?*

To solve this problem, for the first time, we introduce the LLM as a penalty function designer to support safe RL for energy management in ADNs. As shown in Fig. 1, the upstream LLM agent interacts with the ADN operator while the downstream RL agent interacts with the ADN training and test environment. The designed penalty functions bridge the high-level domain knowledge of the LLM agent and the low-level control policy of the RL agent. The LLM agent takes natural language-based requests from the ADN operator as input, then comprehends the energy management problem and safety requirements to generate appropriate penalty functions. And the RL agent updates its control policy with the RL algorithm (soft actor-critic, SAC [29] used in this paper) based on the penalty functions from the LLM agent.

In addition, the penalty functions initially generated by the LLM agent often fail to strictly constrain unsafe actions of the RL agent. To facilitate the adaptive adjustment of the penalty functions to meet the operational safety requirements in ADNs, we propose the RL2 mechanism to iteratively refine the generated penalty functions through multi-round dialogues. In the proposed RL2 mechanism, the RL agent sends its performance back to the LLM agent after each round of training and test, based on which the LLM agent adjusts the penalty functions and their parameters. With its powerful in-context learning capability, the LLM agent can generate appropriate penalty functions that successfully balance the operational cost and safety after several rounds of dialogue. The way in which the LLM agent reinforces its generated content through feedback from the external environment is very similar to RL, so we refer to this mechanism as RL2. Relevant system prompts and reinforcement prompts are also proposed to ensure the LLM performance. The main contributions of this paper are summarized as follows:

1) **LLM as penalty function designer.** For the first time, we introduce LLM as the penalty function designer to assist safe RL for energy management problems in ADNs. The energy management problem is first formulated as an MDP which can be efficiently optimized by the SAC algorithm. Then the LLM agent with embedded domain knowledge is used to comprehend the problem and generate corresponding penalty functions for operational safety constraints, which is added as a penalty term in the final reward received by the RL agent. The idea of designing penalty functions successfully bridges the high-level LLM agent and the low-level RL agent. And the utilization of LLMs greatly reduces the time and cost of the ADN operator.

2) **Mechanism of RL2.** Given the variety of ADN scenarios, it is impossible for the LLM agent to generate penalty functions that satisfy all conditions in a single attempt. Adaptive adaptation of the penalty functions is necessary. Therefore, we propose the RL2 mechanism for the LLM agent to adjust the penalty functions and their parameters through multi-round dialogues using its in-context learning capability. During the multi-round dialogues, the previous penalty functions serve as samples while the RL agent's performance serves as feedback that provides guidance for improving the penalty functions generated in the subsequent rounds.

3) **Designing prompts for the LLM agent.** To assist the LLM agent generate correct content and ensure its performance, we design appropriate system prompts and reinforcement prompts for the LLM agent. The system prompts consist of role description, environment description, task description, output format, penalty function requirements, and rules. And the reinforcement prompts consist of current penalty functions, function evaluation, training results, test results, and chain-of-thought (CoT) [30] suggestions. Extensive experiments on different test cases demonstrate the effectiveness and safety of the proposed method.

The remainder of this paper is organized as follows. Section II formulates the energy management problem in ADNs, and introduces the basic concepts of MDP, safe RL, and LLM used in this paper. Section III details the overall RL2 mechanism, the design of the RL agent, and the design of the LLM agent. Section IV shows the performance of the RL agent and LLM agent on different test cases, which demonstrate the effectiveness of the proposed method. Finally, Section V presents the conclusion and directions for future research.

## II. PRELIMINARIES

In this section, we first formulate the energy management problem in ADNs. Then, the basic concepts about MDP, safe RL, and LLM used in this paper are introduced.

### A. Energy Management Problem Formulation

Considering an ADN equipped with DGs, PVs, and BESSs which can be described as an undirected graph $\mathcal{G}(\mathcal{N}, \mathcal{E})$, where the $\mathcal{N}$ and $\mathcal{E}$ are nodes and branches in the ADN, its active and reactive power injection of node $i$ at time step $t$ can be calculated as:

$$P_{i,t} = P_{i,t}^{DG} + P_{i,t}^{PV} + P_{i,t}^{BESS} - P_{i,t}^{L} \qquad (1)$$

$$Q_{i,t} = Q_{i,t}^{DG} + Q_{i,t}^{PV} - Q_{i,t}^{L} \qquad (2)$$

where $P_{i,t}^{DG}, P_{i,t}^{PV}, P_{i,t}^{BESS}$ represent active power generated by DG, PV, and BESS, respectively; $Q_{i,t}^{DG}, Q_{i,t}^{PV}$ represent reactive power generated by DG and PV, respectively; $P_{i,t}^{L}$ and $Q_{i,t}^{L}$ are the active and reactive load of node $i$ at time step $t$. Then the power flow functions in the ADN are expressed as:

$$P_{i,t} = V_{i,t} \sum_{j \in \mathcal{N}} V_{j,t} (G_{ij} \cos \theta_{ij,t} + B_{ij} \sin \theta_{ij,t}) \qquad (3)$$

$$Q_{i,t} = V_{i,t} \sum_{j \in \mathcal{N}} V_{j,t} (-B_{ij} \cos \theta_{ij,t} + G_{ij} \sin \theta_{ij,t}) \qquad (4)$$

where $V_{i,t}$ is the voltage magnitude at node $i$ at time step $t$; $\theta_{ij,t}$ is the voltage phase difference between node $i$ and $j$ at time step $t$; $G_{ij}$ and $B_{ij}$ are the real and imaginary parts of the corresponding element $Y_{ij} = G_{ij} + jB_{ij}$ in the admittance matrix of the ADN.

By setting energy management control signals for $P_{i,t}^{DG}, P_{i,t}^{BESS}, Q_{i,t}^{DG}, Q_{i,t}^{PV}$, the ADN operator aims to minimize the total operational cost:

$$\min \sum_{t=0}^{T} [\sum_{i \in \mathcal{N}} C_i^{DG}(t) + \sum_{i \in \mathcal{N}} C_i^{BESS}(t) + C_0(t)] \qquad (5)$$

where $T$ is the length of the control process; $C_i^{DG}(t)$ is the generation cost of the DG at node $i$ at time step $t$; $C_i^{BESS}(t)$ is the charge or discharge cost of the BESS at node $i$ at time step $t$; $C_0(t)$ is the cost of buying electricity from the upper grid. These operational costs can be calculated as:

$$C_i^{DG}(t) = \rho_i^{DG} P_{i,t}^{DG} \qquad (6)$$

$$C_i^{BESS}(t) = \begin{cases} \rho_{i,dis}^{BESS} P_{i,t}^{BESS}, & P_{i,t}^{BESS} \geq 0 \\ -\rho_{i,ch}^{BESS} P_{i,t}^{BESS}, & P_{i,t}^{BESS} < 0 \end{cases} \qquad (7)$$

$$C_0(t) = \begin{cases} \rho_{buy,t} P_{0,t}, & P_{0,t} \geq 0 \\ \rho_{sell,t} P_{0,t}, & P_{0,t} < 0 \end{cases} \qquad (8)$$

where $\rho_i^{DG}$ is the cost coefficient of the DG at node $i$; $\rho_{i,ch}^{BESS}$ and $\rho_{i,dis}^{BESS}$ are the charge and discharge cost coefficients of the BESS at node $i$; $\rho_{buy,t}$ and $\rho_{sell,t}$ are prices for buying or selling electricity at time step $t$; $P_{0,t}$ is the active power supplied by the upper grid at time step $t$.

Apart from the operational costs, there are also several constraints that need to be considered during optimization, including equipment constraints (9)-(15) and operational safety constraints (16)-(17):

Equipment constraints of the DGs are presented in (9)-(11), in which $P_{i,min}^{DG}, P_{i,max}^{DG}$ are limitations of active power generation of DG at node $i$; $Q_{i,min}^{DG}, Q_{i,max}^{DG}$ are limitations of reactive power generation of DG at node $i$; $R_{i,down}$ and $R_{i,up}$ are the maximum ramping rate.

$$P_{i,min}^{DG} \leq P_{i,t}^{DG} \leq P_{i,max}^{DG}, \forall i \in \mathcal{N} \qquad (9)$$

$$Q_{i,min}^{DG} \leq Q_{i,t}^{DG} \leq Q_{i,max}^{DG}, \forall i \in \mathcal{N} \qquad (10)$$

$$-R_{i,down} \leq P_{i,t}^{DG} - P_{i,t-1}^{DG} \leq R_{i,up}, \forall i \in \mathcal{N} \qquad (11)$$

Equipment constraints of the PVs are presented in (12), in which $S_{i,max}^{PV}$ is the installed capacity of the PV at node $i$.

$$\left(P_{i,t}^{PV}\right)^2 + \left(Q_{i,t}^{PV}\right)^2 = \left(S_{i,t}^{PV}\right)^2 \leq \left(S_{i,max}^{PV}\right)^2, \forall i \in \mathcal{N} \qquad (12)$$

Equipment constraints of the BESSs are presented in (13)-(15), in which $P_{i,min}^{BESS}, P_{i,max}^{BESS}$ are limitations of active power generation of BESS at node $i$; $SOC_{i,min}^{BESS}, SOC_{i,max}^{BESS}$ are limitations of state of charge of BESS at node $i$; $SOC_{i,t}^{BESS}$ is the current state of charge; $\Delta t$ is the interval between two timesteps; $\eta$ is the charge and discharge efficiency.

$$P_{i,min}^{BESS} \leq P_{i,t}^{BESS} \leq P_{i,max}^{BESS}, \forall i \in \mathcal{N} \qquad (13)$$

$$SOC_{i,min}^{BESS} \leq SOC_{i,t}^{BESS} \leq SOC_{i,max}^{BESS}, \forall i \in \mathcal{N} \qquad (14)$$

$$SOC_{i,t}^{BESS} = \begin{cases} SOC_{i,t-1}^{BESS} - \frac{P_{i,t}^{BESS} \Delta t}{\eta}, & P_{i,t}^{BESS} \geq 0 \\ SOC_{i,t-1}^{BESS} - \eta P_{i,t}^{BESS} \Delta t, & P_{i,t}^{BESS} < 0 \end{cases}, \forall i \in \mathcal{N} \qquad (15)$$

Operational constraints in the ADN mainly include voltage constraints (16) and branch power constraints (17), in which the voltage magnitudes must not exceed the upper or lower limitations, and the branch power must not exceed the capacity.

$$V_{min} \leq V_{i,t} \leq V_{max}, \forall i \in \mathcal{N} \qquad (16)$$

$$\left(P_{ij,t}^{brch}\right)^2 + \left(Q_{ij,t}^{brch}\right)^2 = \left(S_{ij,t}^{brch}\right)^2 \leq \left(S_{max}^{brch}\right)^2, \forall ij \in \mathcal{E} \qquad (17)$$

where $V_{min}, V_{max}$ are voltage limitations; $P_{ij,t}^{brch}$ and $Q_{ij,t}^{brch}$ are the active and reactive branch power in branch $ij$ at time step $t$; $S_{max}^{brch}$ is the branch capacity.

According to the definition of RL, when the RL agent interacts with the training and test environment, equipment constraints can be easily satisfied by properly setting the action space of the MDP, while the operational safety constraints should be treated as penalty terms added in the final reward received by the RL agent. Since the operational safety constraints (16)-(17) cannot be directly interpreted as information that can be learned by the RL agent, we aim to instruct the LLM agent to comprehend the environment and generate appropriate penalty functions. These two different types of constraints need to be correctly comprehended and interpreted by the LLM agent.

### B. Markov Decision Process and Safe Reinforcement Learning

In order to formalize sequential decision process of the RL agent and solve the energy management problem with SAC algorithm, we briefly introduce MDP in this subsection. An MDP is formulated as $(\mathcal{S}, \mathcal{A}, p, \mathcal{R}, [\mathcal{C}_n]_N, \gamma)$, where $\mathcal{S}$ and $\mathcal{A}$ are the state space and action space of the environment; $p$ is the state transition function; $\mathcal{R}$ is the reward function; $\mathcal{C}_1, \ldots, \mathcal{C}_N$ are the penalty functions; $\gamma \in [0,1)$ is the discount factor for future rewards.

At time step $t$, the RL agent receives current state $s_t \in \mathcal{S}$ of the environment, then executes an action $a_t \in \mathcal{A}$ based on its control policy $\pi: \mathcal{S} \times \mathcal{A} \to [0, \infty)$, i.e., $a_t \sim \pi(\cdot \mid s_t)$. Afterwards, the environment transfers to next state $s_{t+1} \in \mathcal{S}$ according to $p$ and sends the reward $r_t = \mathcal{R}(s_t, a_t)$ and penalties $c_{1,t} = \mathcal{C}_1(s_t, a_t), \ldots, c_{N,t} = \mathcal{C}_N(s_t, a_t)$ as feedback to the RL agent, based on which the RL agent updates its policy to maximize the reward while minimizing the penalties.

In safe RL, $c_{1,t}, \ldots, c_{N,t}$ are usually added as penalty terms into $r_t$ so that the final reward received by the RL agent is $r'_t = \beta_r r_t - \sum_{n=1}^N \beta_{n,c} c_{n,t}$. Here, $\beta_r$ is the weight coefficient for the reward $r_t$ and $\beta_{n,c}$ is the weight coefficient for the $n$th penalty $c_{n,t}$. Then the objective of the RL agent is to maximize its expected discounted cumulative reward:

$$\max_{\pi} J(\pi) = \mathbb{E}_{p,\pi}[\sum_{t=0}^T \gamma^t r'_t] \qquad (18)$$

To better illustrate the SAC algorithm used in this paper, we define the action value function $Q^\pi(s, a)$ here, which calculates the expected discounted cumulative reward after taking action $a$ at state $s$ under the policy $\pi$:

$$Q^\pi(s, a) := \mathbb{E}_{p,\pi}[\sum_{t=0}^T \gamma^t r'_t \mid s_0 = s, a_0 = a] \qquad (19)$$

### C. Large Language Model

In recent years, LLMs have emerged as a pivotal breakthrough in the field of artificial intelligence, particularly in natural language processing, computer vision, and multimodal applications. These LLMs, such as ChatGPT [31], characterized by their millions to trillions of parameter sizes, have revolutionized the way machines understand and generate human-like content. An extensive corpus of datasets from diverse sources such as Internet, books, and articles enables the LLMs to develop a deep understanding of language, generate coherent content, perform complex reasoning, and engage in meaningful dialogue. Once trained, LLMs exhibit a strong ability to predict the next word in a sequence given the preceding context. The probability of the next word is modelled as:

$$\Pr(\omega_1, \ldots, \omega_M \mid context) = \prod_{m=1}^M \Pr(\omega_m \mid \omega_{1:m-1}, context) \qquad (20)$$

where $\omega_1, \ldots, \omega_M$ is the generated sequence; $context$ is the given input. In our proposed RL2, the $context$ includes not only the prompts, but also the previously designed penalty functions from the LLM agent itself.

During inference, the model generates content by sampling from this probability distribution or selecting the word with the highest probability. This process can be repeated to generate sequences of arbitrary length, making LLMs highly effective for tasks such as text completion, story generation, and dialogue systems. In addition, techniques such as unsupervised pre-training, supervised fine-tuning, and RL are commonly employed to improve the LLMs' performance in specific tasks. However, these techniques are beyond the scope of this paper, which focuses on utilizing existing LLMs to assist safe RL for ADN operators in energy management problems.

Compared to traditional artificial intelligence, LLMs have demonstrated powerful capabilities across multiple dimensions, particularly in the areas of comprehension, reasoning, and in-context learning. Comprehension refers to the LLMs' ability to understand and interpret complex linguistic structures and contexts. LLMs achieve this through their transformer structures, allowing them to capture complicated patterns and dependencies in the input text, which are necessary for ADN operators whose input may include some colloquial or vague expressions. Reasoning is the ability of LLMs to perform logical and inferential thinking based on input, which involves drawing conclusions, solving problems, and making predictions. LLMs can reason about both explicit and implicit information, allowing them to answer questions that require multi-step logical reasoning. Reasoning ability is also critical for safe RL in ADNs since the LLMs need to associate the operational safety requirements with corresponding penalty functions. The in-context learning ability is a unique feature of LLMs that allows them to adapt and improve their performance based on the given context. Unlike traditional models that require explicit training on specific tasks, LLMs can leverage the information given in the prompts and previous dialogues to generate appropriate responses. This dynamic adaptation of in-context learning ability is particularly useful in scenarios where the LLMs may encounter new ADN environments and other different types of operational safety requirements.

### III. Methods

In this section, we first formulate the MPD for the energy management problem in ADNs so that it can be effectively optimized by SAC algorithm. Then the overall RL2 mechanism is described in detail. Within the RL2 mechanism, design of the RL agent as well as the LLM agent are then introduced.

### A. Markov Decision Process Formulation for Energy Management

Definitions of state space, action space, reward function, and penalty functions for the energy management problem are designed as follows.

1) *State Space*: The state $s \in \mathcal{S}$ of the MDP for energy management problem is based on the measurements and states of the equipment in the ADN. In this paper, we assume all the nodes and devices are observable, then the state can be defined as a vector:

$$s = (P_i^{DG}, P_i^{PV}, P_i^{BESS}, P_i^L, Q_i^{DG}, Q_i^{PV}, Q_i^L, V_i, SOC_i^{BESS}), \forall i \in \mathcal{N} \tag{21}$$

2) *Action Space*: The action $a \in \mathcal{A}$ of the MDP for energy management problem is based on the controllable devices in the ADN. In this paper, we assume all the DGs, PVs, and BESSs within the ADN are controllable for the ADN operator, then the action can also be defined as a vector:

$$a = (P^{DG}, P^{BESS}, Q^{DG}, Q^{PV}), \forall i \in \mathcal{N} \tag{22}$$

3) *Reward Function*: Designing the reward function of the MDP for energy management problems is straightforward, as it is consistent with the objective function. In other words, maximizing the RL agent's reward is equivalent to minimizing the operational cost of the ADN. Therefore, the reward function is defined as:

$$r_t = \mathcal{R}(s_t, a_t) = -\left[ \sum_{i \in \mathcal{N}} C_i^{DG}(t) + \sum_{i \in \mathcal{N}} C_i^{BESS}(t) + C_0(t) \right] \tag{23}$$

4) *Penalty Functions*: Although the operational safety requirements of the ADN operator can be easily specified in natural language, they are difficult for the RL agent to understand, which complicates the task of designing penalty functions for the MDP. Hence, it is necessary to employ the LLM agent to comprehend different types of constraints and generate corresponding penalty functions. Under this circumstance, the designed penalty functions bridge the gap between high-level domain knowledge of the LLM agent and low-level control policy of the RL agent. For convenience, we denote the penalty function for voltage constraints (16) as $c_{V,t} = \mathcal{C}_V(V_{i,t}), \forall i \in \mathcal{N}$ and the penalty function for branch power constraints (17) as $c_{brch,t} = \mathcal{C}_{brch}(S_{ij,t}^{brch}), \forall ij \in \mathcal{E}$.

5) *Final Reward*: As we mentioned before, the $c_{V,t}$ and $c_{brch,t}$ are added as penalty terms into $r_t$ to calculate the final reward $r_t'$:

$$r_t' = \beta_r \mathcal{R}(s_t, a_t) - \frac{\sum_{\forall i \in \mathcal{N}} \beta_V \mathcal{C}_V(V_{i,t})}{|\mathcal{N}|}$$
$$- \frac{\sum_{\forall ij \in \mathcal{E}} \beta_{brch} \mathcal{C}_{brch}(S_{ij,t}^{brch})}{|\mathcal{E}|} \tag{24}$$

It should be noted that the coefficient $\beta_r$ for the reward function $\mathcal{R}(s_t, a_t)$ is explicitly assigned by the ADN operator while the coefficients $\beta_V, \beta_{brch}$ for the penalties $\mathcal{C}_V(V_{i,t}), \mathcal{C}_{brch}(S_{ij,t}^{brch})$ are implicitly assigned by the LLM agent, i.e., $\beta_V \mathcal{C}_V(V_{i,t})$ and $\beta_{brch} \mathcal{C}_{brch}(S_{ij,t}^{brch})$ are the complete penalty functions the LLM agent designs.

### B. Overall RL2 Mechanism

Besides employing LLMs for designing penalty functions, we also propose an RL2 mechanism to enable the LLM agent to adaptively adjust penalty functions in accordance with the operational safety requirements in the ADN. Utilizing the in-context learning capability of LLMs, the RL2 mechanism instructs the LLM agent to iteratively adjust the functions' patterns and parameters through multi-round dialogues based on feedback from the RL agent.

In the proposed RL2 mechanism, the LLM agent is first informed about the energy management problem of the ADN and the requirements for penalty functions through system prompts. After the generation of the penalty functions, these functions are then set up in the ADN training and test environments so that the RL agent can perceive the safe operational boundaries of the ADN. The next step is the RL training based on the SAC algorithm to obtain training results and RL policy $\pi$. A typical day of the ADN is then selected to test the RL policy's performance. Once the test is completed, training results, test results, and current penalty functions are embedded into the reinforcement prompts and sent back to the LLM agent as feedback, based on which the LLM agent will redesign the penalty functions to improve the safety performance or balance the operational cost. The entire process requires little intervention from the ADN operator, significantly improving the safe RL efficiency in energy management problems.

Overall RL2 mechanism is listed in Algorithm 1. Details of the RL agent and LLM agent are described in the following subsections.

---

**Algorithm 1**: RL2

---

**Initialize**: LLM agent, RL agent, ADN training and test environments, system prompts, reinforcement prompts, $history = []$, iteration limit $K$, current iteration $k = 0$
**Output**: RL agent control policy, final penalty functions

1  **while** $k < K$ **do**
2    **if** $k = 0$ **do**
     *// Establish system prompts*
3      Append $history \leftarrow system\ prompts$
4    **end if**
   *// LLM design penalty functions*
5    $penalty\ functions \sim LLM(history)$
6    Append $history \leftarrow penalty\ functions$
   *// RL training*
7    Set $ADN \leftarrow penalty\ functions$
8    $\pi, training\ results \sim RL\_training(ADN)$
   *// RL test*
9    $test\ results \sim RL\_test(ADN | \pi)$
   *// Embed results in reinforcement prompts*
10   Embed $reinforcement\ prompts \leftarrow training\ results, test\ results, penalty\ functions$
   *// Establish reinforcement prompts*
11   Append $history \leftarrow reinforcement\ prompts$
12   $k = k + 1$
13  **end while**
14  **return** $\pi, penalty\ functions$

---

### C. Soft Actor-Critic for RL Agent

In this paper, the SAC algorithm is utilized to formulate the RL agent and optimize the MDP for the energy management problem, which exhibits excellent learning and exploration abilities. In SAC, the policy $\pi$ of the RL agent is approximated by a policy network $\pi_\theta$ with parameters $\theta$ so that the action $a$ can be reparameterized as:

$$a_\theta(s,\xi) = \tanh(\mu_\theta(s) + \sigma_\theta(s) \odot \xi), \ \xi \sim \mathcal{N}(0, I) \quad (25)$$

where $\mu_\theta, \sigma_\theta$ approximate the mean value and standard deviation of the action distribution, respectively. The action value function $Q^\pi(s, a)$ is also approximated by a value network $Q_\varphi$ with parameters $\varphi$.

During the interaction with the training environment, the RL agent with SAC algorithm stores each transition information $(s_t, a_t, r'_t, s_{t+1})$ in a replay buffer, from which the RL agent randomly samples batches of data $\mathcal{B}$ to update the networks. Loss function of the value network $Q_\varphi$ is calculated as:

$$\mathcal{L}(\varphi) = \frac{1}{|\mathcal{B}|} \Sigma_{s_t, a_t, r'_t, s_{t+1}} \left[ \left( Q_\varphi(s_t, a_t) - y_t \right)^2 \right] \quad (26)$$

Here, $y_t$ is the target value for $Q_\varphi(s_t, a_t)$, which is expressed as:

$$y_t = r'_t + \gamma \left[ Q_{\varphi'}(s_{t+1}, a_{t+1}) - \alpha \log \pi_\theta(a_{t+1}|s_{t+1}) \right] \quad (27)$$

where $Q_{\varphi'}$ is the target value network introduced to stabilize the training process, whose parameters $\varphi'$ are gradually updated by parameters $\varphi$; $a_{t+1}$ is an action sampled from the distribution $\pi_\theta(s_{t+1})$; $\mathbb{E}[-\log \pi_\theta(\cdot|s_{t+1})]$ is the entropy term measuring stochasticity of the policy added by the SAC algorithm to encourage exploration of the RL agent during training, whose weight coefficient is $\alpha$.

Loss function of the policy network $\pi_\theta$ is calculated as:

$$\mathcal{L}(\theta) = \frac{1}{|\mathcal{B}|} \Sigma_{s_t} \left[ -Q_\varphi(s_t, a_t) + \alpha \log \pi_\theta(a_t|s_t) \right] \quad (28)$$

where $a_t$ is an action sampled from the distribution $\pi_\theta(s_t)$. After the calculation of the loss functions, the policy network $\pi_\theta$ and value network $Q_\varphi$ of the RL agent are updated using gradient descent:

$$\theta \leftarrow \theta - \lambda_\theta \nabla_\theta \mathcal{L}(\theta) \quad (29)$$
$$\varphi \leftarrow \varphi - \lambda_\varphi \nabla_\varphi \mathcal{L}(\varphi) \quad (30)$$

where $\lambda_\theta$ and $\lambda_\varphi$ are learning rates of corresponding parameters. And the target value network $Q_{\varphi'}$ is updated as:

$$\varphi' \leftarrow \lambda_{\varphi'} \varphi + (1 - \lambda_{\varphi'}) \varphi' \quad (31)$$

where $\lambda_{\varphi'}$ is the updating rate of $\varphi'$.

### D. Prompts Design for LLM Agent

As can be seen from the SAC algorithm, performance of the RL agent relies heavily on the received final reward, in which the penalty functions are designed by the LLM agent. In order to facilitate the design and adjustment of the penalty functions from the LLM agent, appropriate prompts should be crafted to support its content generation. Based on our proposed RL2 mechanism, our prompts consist of two parts, system prompts and reinforcement prompts. System prompts are tasked with describing the energy management problem and supporting the design of the penalty functions, and reinforcement prompts focus on providing performance feedback from the RL agent and assisting in the adjustment of the penalty functions. Details of the prompts are as follows, and corresponding components are shown in Fig. 2.

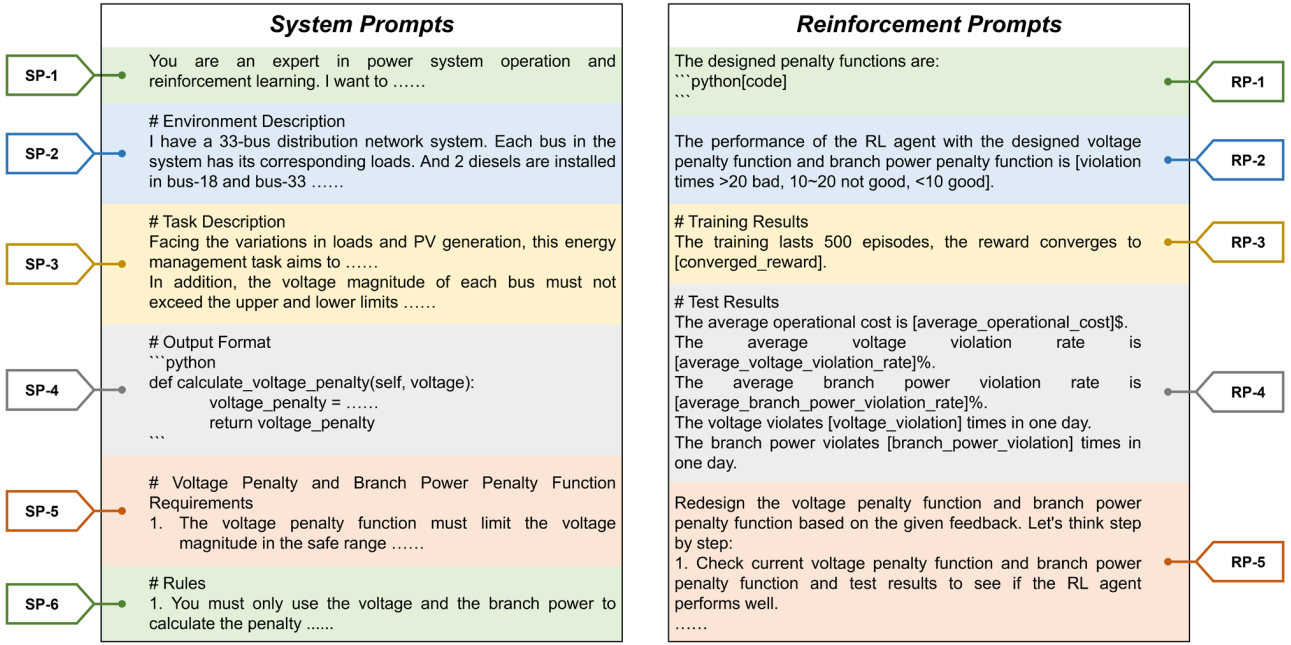

**Fig. 2.** Illustration of the proposed system prompts and reinforcement prompts.

1) *System prompts*: Before the dialogues start, the system prompts are established for the LLM agent to specify its role, describe the energy management problem, environment, and task, and inform the requirements for penalty functions. The system prompts include six components:

(SP-1) Role description: This component describes the role of the LLM agent, which is an expert in power system operation and RL in our problem, helping the LLM agent to extract relevant domain knowledge from its embedded world knowledge.

(SP-2) Environment description: This component describes the energy management problem and the ADN environment for the RL agent, which helps the LLM agent adapt to different scenarios.

(SP-3) Task description: This component describes the task of the LLM agent, which is to design penalty functions for the RL agent to ensure safety constraints of the ADN are satisfied. Parameters concerning operational safety requirements, including the lower limit $V_{min}$ and upper limit $V_{max}$ of the voltage magnitude, and the branch capacity $S_{max}^{brch}$, are also detailed in this component, which is necessary for the design of penalty functions.

(SP-4) Output format: This component describes the output format for the designed penalty functions $\beta_V \mathcal{C}_V(V_{i,t})$ and $\beta_{brch} \mathcal{C}_{brch}(S_{ij,t}^{brch})$. In our problem, we instruct the LLM agent to generate two Python functions, `calculate_voltage_penalty` whose input is $V_{i,t}$ and `calculate_branch_power_penalty` whose input is $S_{ij,t}^{brch}$.

(SP-5) Penalty functions requirements: This component describes the requirements of the penalty functions. In safe RL for energy management, the penalty functions should limit voltage magnitudes within the lower limit and upper limit, and limit branch powers below the branch capacity. To avoid hallucination of the LLM agent, we also require the pattern of the penalty functions should be simple enough.

(SP-6) Rules: This component describes some rules that need to be emphasized before the content generation.

*2) Reinforcement prompts*: As can be seen from the proposed RL2 mechanism, the reinforcement prompts are fed back to the LLM agent each time it designs penalty functions, serving to inform the performance of the penalty functions and suggest directions for future adjustment. The reinforcement prompts include five components:

(RP-1) Current penalty functions: This component outlines current penalty functions, so that the LLM agent can associate the RL agent's performance with corresponding penalty functions.

(RP-2) Functions evaluation: When training is completed, we will evaluate the RL agent on a typical day of the ADN, which will serve as the criterion for the designed penalty functions. This component summarizes in one sentence whether the designed functions are effective or not, indicating the preference of the ADN operator. In our energy management problem, we set a threshold for the violation times on the typical day. If the violation times on the typical day exceed this threshold, we label the designed functions as "bad", otherwise they are labeled as "good".

(RP-3) Training results: This component describes training results of the RL agent, including the converged reward.

(RP-4) Test results: This component describes test results of the RL agent on the typical day, including the operational cost, voltage violation rate, branch power violation rate, voltage violation times, and branch power violation times.

(RP-5) CoT suggestions: Finally, we request the LLM agent to redesign the penalty functions based on the provided feedback and offer some suggestions, which are based on the CoT. CoT technique involves breaking down complex reasoning tasks into a sequence of simpler intermediate steps to improve the accuracy of the LLMs. By guiding the LLM agent through a structured reasoning process, CoT enhances logical coherence and reliability of the LLM agent. It should be noted that, as shown in Fig. 2, our CoT suggestions require little domain knowledge in RL and power system operation, which is suitable for the application of emerging ADN operators.

*3) Code self-verification*: Considering the limitations of LLMs, an essential step after each generation of the penalty functions is the code self-verification. The self-verification consists of two aspects: the applicability of the code and the rationality of the penalty functions. Our self-verification leverages the code testing capability of the LLM agent, which has been embedded in the majority of existing LLMs. Since we request the LLM agent to generate Python functions in the system prompts, the applicability of the code can be easily tested by direct execution. As for the rationality of the penalty functions, we can input a list of voltage magnitudes and branch powers into the penalty functions to see if the function values increase as the violations become more severe. If the code self-verification does not pass, we can simply ask the LLM agent to regenerate new penalty functions until the self-verification is successful.

## IV. NUMERICAL STUDY

In this section, to validate the effectiveness of the proposed method, several numerical simulations are conducted on IEEE 33-bus [32] and 69-bus [33] distribution networks. Control interval is 15 minutes and the length of the control process is 96 steps. A steady state ADN RL training and test environment is built under the scheme of toolkit Gym [34] using power flow functions.

In 33-bus system, 2 DGs are located at bus-18 and bus-33, 2 PVs are located at bus-22 and bus-25, and 2 BESSs are located at bus-21 and bus-24. In 69-bus system, 2 DGs are located at bus-18 and bus-58, 2 PVs are located at bus-35 and bus-46, and 2 BESSs are located at bus-34 and bus-45. The branch capacities of these two cases are 5.0 MVA and 4.7 MVA, respectively. Voltage limitations are $[0.95, 1.05] p.u.$.

### A. Proposed Method Setup

In this subsection, we setup the LLM agent in our proposed method with one of the most powerful existing LLMs qwen-max[1] and the RL agent with SAC algorithm. All of the RL agents are implemented in Python with the deep learning framework Pytorch. Experiments are run on a computer with a 2.3GHz Intel Core i7-10875H CPU and 16GB RAM. Considering the randomness of the LLM agent and SAC algorithm, for each group of experiments, we test with 5 independent random seeds. Hyperparameters for the LLM agent and the RL agent are listed in Table I and Table II.

---

[1] Version of qwen-max in this paper is "2024-09-19".

In addition, a mixed-integer second-order cone programming (MISOCP) based on DistFlow [35] of the ADN is also implemented for comparison, which could be considered as a theoretically optimal result.

TABLE I
HYPERPARAMETERS OF THE LLM AGENT

| Parameters | Value |
|---|---|
| Model | qwen-max |
| Temperature | 0.8 |
| Top-p | 0.8 |

TABLE II
HYPERPARAMETERS OF THE RL AGENT

| Parameters | Values |
|---|---|
| Optimizer | Adam |
| Non-linearity | ReLU |
| Replay buffer size | $1 \times 10^4$ |
| Batch size | 256 |
| Hidden layers | 3 |
| Hidden units | 256 |
| Value update frequency | 1 |
| Policy update frequency | 2 |
| $\gamma$ | 0.99 |
| $\lambda_\varphi$ | 1.0e-3→1.0e-4 (Exponential decay) |
| $\lambda_{\varphi'}$ | 5.0e-3 |
| $\lambda_\theta$ | 1.0e-3→1.0e-4 (Exponential decay) |
| $\alpha$ | 0.04 |

### B. Performance of LLM Agent

After the method setup, we establish the system prompts proposed in Section III for the LLM agent and require it to generate penalty functions for corresponding operational safety requirements. After each group of training and test is completed, we embed the performance of the RL agent into the reinforcement prompts and send them back to the LLM agent to adjust the penalty functions.

In order to better visualize the process of designing and adjusting penalty functions by the LLM agent, we illustrate an example in Fig. 3, with key parts highlighted. Detailed prompts, designed functions, and dialogue process of each group of experiments are listed in the supplementary file [36], which also presents the progressive performance of the RL agent.

The first observation from Fig. 3 is that the LLM agent can correctly extract safety-related parameters from the input with our proposed system prompts, such as $V_{min} = 0.95 p.u.$, $V_{max} = 1.05 p.u.$, and $S_{max}^{brch} = 5.0 MVA$ in this case. Even at the beginning, the LLM agent can design correct penalty functions for different types of safety constraints, i.e., voltage constraints and branch power constraints, and both penalty functions are valid, with their values growing as the violations increase. These observations indicate the LLM agent is capable of comprehending the energy management problem in the ADN and its operational safety requirements, and then generating appropriate penalty functions. This feature can be utilized in safe RL to guide the RL agent towards safer policies, which greatly reduces the time and expenses for the ADN operator.

Additionally, Fig. 3 also shows how the designed penalty functions are adjusted. Usually, the initially designed penalty functions (Function initialization stage) are insufficient to meet the safety requirements of the ADN. Subsequently, the LLM agent starts a search with some randomness in the function space (Exploration stage), with the goal of enhancing safety. Once the LLM agent identifies penalty functions that significantly improve safety of the RL agent, the penalty function patterns are then determined (Pattern determination stage), for example, piecewise polynomial function in this experiment. Finally, to achieve a balance between the operational cost and safety, the LLM agent adjusts penalty function parameters accordingly (Parameter adjustment stage).

It is interesting to note that the above four stages of the LLM agent are very similar to RL, and the process of adjusting the penalty functions by the LLM agent remarkably enhances safety of the RL agent, which demonstrates the effectiveness of our proposed RL2 mechanism.

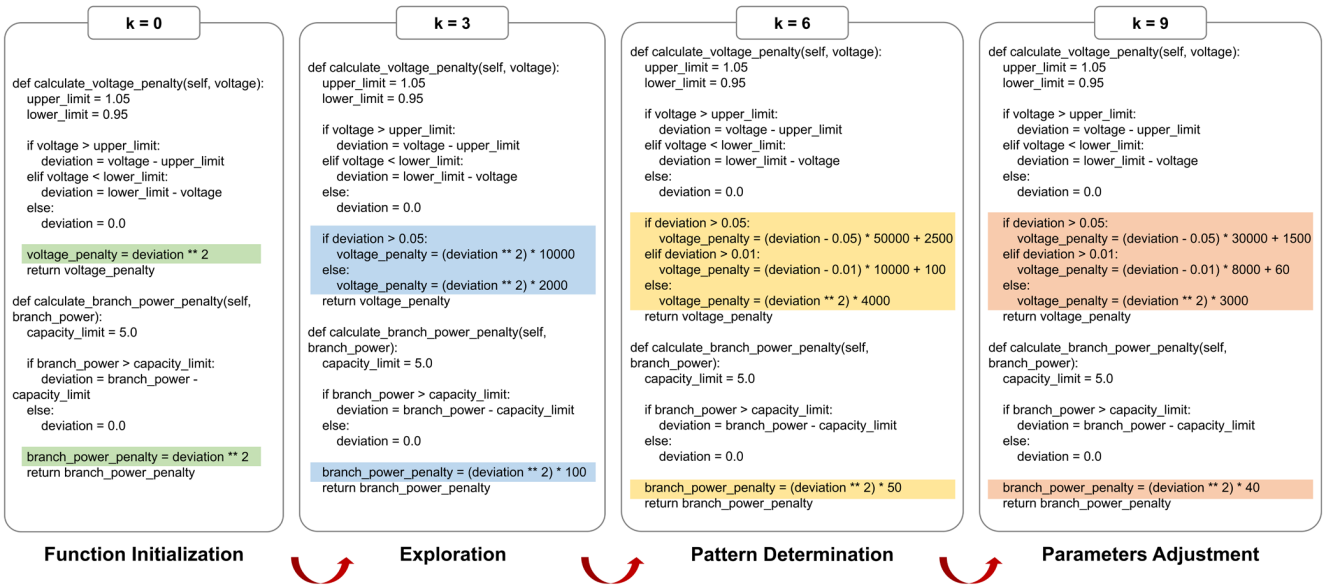

**Fig. 3.** Illustration of one group of designed penalty functions.

## C. Performance of RL Agent

To validate the effectiveness of the SAC algorithm for energy management problem and the effectiveness of the proposed RL2 mechanism, we compare the training and test results of the RL agent under three different penalty functions, i.e., initial penalty functions designed by the LLM agent (Init.), penalty functions in the middle of the adjustment (Intermediate penalty functions, Mid.), and the final penalty functions designed by the LLM agent (Fin.). Also, to quantify the safety performance of the RL agent, we define two safety indexes called voltage violation rate (VVR) and branch power violation rate (BVR), which are calculated as:

$$VVR_t = \frac{\sum_{i \in \mathcal{N}} [\mathbb{1}(V_{i,t} > V_{max}) + \mathbb{1}(V_{i,t} < V_{min})]}{|\mathcal{N}|} \tag{32}$$

$$BVR_t = \frac{\sum_{ij \in \mathcal{E}} [\mathbb{1}(S_{ij,t}^{brch} > S_{max}^{brch})]}{|\mathcal{E}|} \tag{33}$$

where $\mathbb{1}(\cdot)$ is the indicator function. Since BVR fluctuates during training and converges near zero when the training process is completed, we utilize step average VVR to reflect safety in the following figures and tables. Complete results can be found in the supplementary file [36].

In Fig. 4-5, we present the training process of the RL agent in 33-bus system and 69-bus system, respectively. Training lasts 500 episodes for each agent. Mean values and error bounds across the five random seeds are depicted as solid lines and shading areas in the figures.

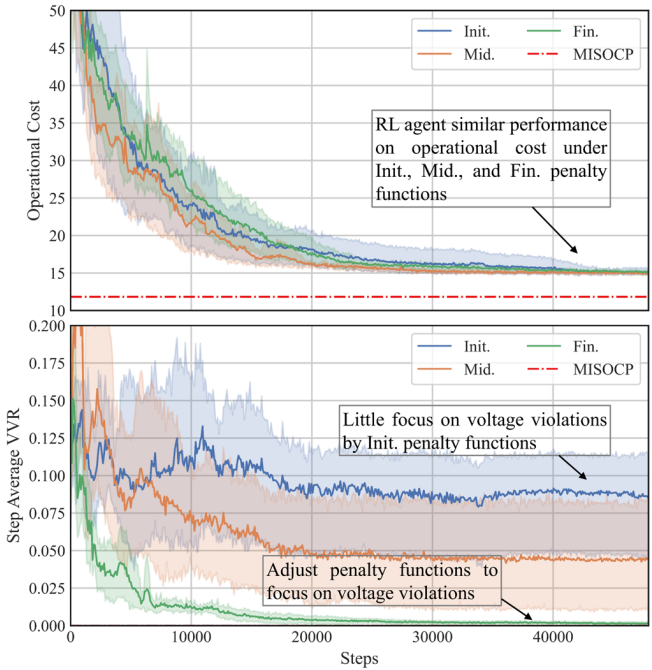

**Fig. 4.** Training process of the RL agent under different penalty functions in 33-bus system.

As can be seen from Fig. 4, all of the RL agents converge to a lower operational cost near the optimal results, which demonstrates the RL-based method in energy management problems. On the other hand, since the Init. penalty functions pay little attention to the RL agent's safety, many violations occur during and after the training. With the adjustment of the penalty functions, the RL agent's safety progressively improves, and VVR under Mid. penalty functions is much lower compared to the Init. penalty functions. Finally, the Fin. penalty functions successfully assist the RL agent to converge to a much safer policy, with little sacrifice on the operational cost.

In 69-bus system, although the operational cost of the RL agent under Init. penalty functions is very close to the optimal result, its safety performance is poor, and VVR even exhibits an upward trend during the training process. While safety performance of the Mid. penalty functions demonstrates a certain level of improvement, it is still unsatisfactory. Also, performance of the Mid. penalty functions varies greatly across the random seeds, as reflected by the large error bounds area. Fin. penalty functions manage to minimize the violations, at the cost of a little higher operational cost, which not only demonstrates the effectiveness of RL2 but also the ability of the LLM agent to balance operational cost and safety requirements.

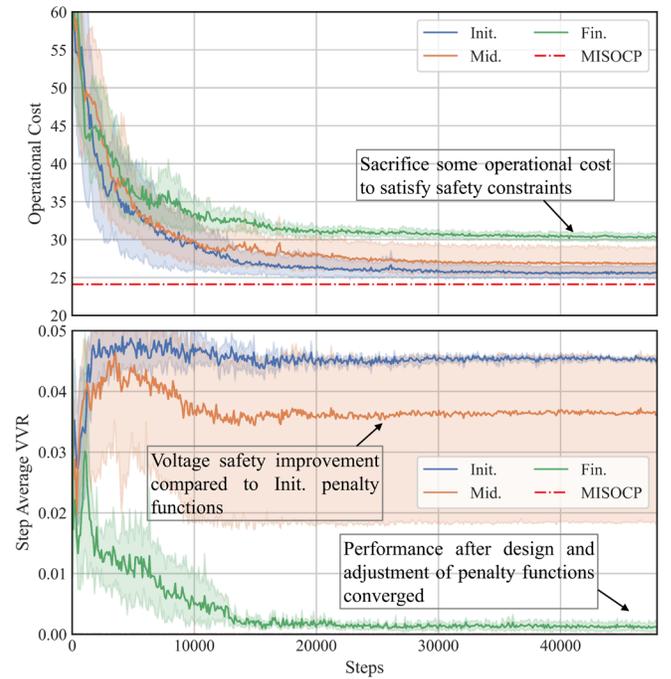

**Fig. 5.** Training process of the RL agent under different penalty functions in 69-bus system.

After training, we list the test results of the RL agents on the typical day in Table III. "Mean" and "Std." in Table III denote mean values and standard deviations across five random seeds. Best results in each column are marked in bold.

TABLE III
TEST RESULTS UNDER DIFFERENT PENALTY FUNCTIONS

| Sys. | Fun. | Operational Cost | | VVR | |
|------|------|------|------|------|------|
| | | Mean | Std. | Mean | Std. |
| 33-bus | Init. | 15.04 | 4.82e-01 | 8.50e-02 | 4.09e-02 |
| | Mid. | **14.87** | **2.06e-01** | 4.39e-02 | 4.11e-02 |
| | Fin. | 15.12 | 2.97e-01 | **1.39e-03** | **1.32e-03** |
| | MISOCP | 11.83 | - | 0.0 | - |
| 69-bus | Init. | **25.73** | 1.02 | 4.53e-02 | **3.02e-04** |
| | Mid. | 26.99 | 2.17 | 3.63e-02 | 1.80e-02 |
| | Fin. | 30.38 | **4.96e-01** | **1.12e-03** | 7.06e-04 |
| | MISOCP | 24.11 | - | 0.0 | - |

During test, RL agents under Fin. penalty functions remain the safest ones, which is consistent with the training results. The improvement in the RL agents' performance is also evident, which verifies the effectiveness of the proposed RL2 mechanism again.

## V. CONCLUSION

Utilizing safe RL method to optimize energy management problems in ADNs requires extensive domain knowledge in RL and power system operation, which heavily depends on the ADN operators' experiences. To improve the efficiency of automatically safe RL, we introduce LLMs to comprehend the energy management problem and design penalty functions for corresponding operational safety constraints. In addition, an RL2 mechanism is also proposed to adaptively refine the designed functions, in which the penalty functions are iteratively adjusted through multi-round dialogues based on feedback from the RL agent. Comprehensive results demonstrate effectiveness and safety of the proposed method.

In future work, other approaches to combining LLMs with RL will be an interesting topic. With the accelerated development of LLMs, other applications in the power system can also be investigated.